\documentclass[reprint, aps,prd,twocolumn,showpacs,showkeys, 10pt, longbibliography, nolinenumbers, floatfix]{revtex4-2}

\usepackage{amsmath}
\usepackage{graphicx}
\usepackage{float}
\usepackage{dcolumn}
\usepackage{multirow}
\usepackage{natbib}
\usepackage{changes}
\usepackage[colorlinks = true,linkcolor = blue,urlcolor  = blue,citecolor = blue,anchorcolor = blue]{hyperref}
\usepackage{enumerate}

\begin{document}

\title{Signatures of quark deconfinement through the $r$-modes of twin stars}

\author{P. Laskos-Patkos}
\email{plaskos@physics.auth.gr}

\author{Ch.C. Moustakidis}
\email{moustaki@auth.gr}

\affiliation{Department of Theoretical Physics, Aristotle University of Thessaloniki, 54124 Thessaloniki, Greece}

\begin{abstract}
The observation and distinction of two compact stars with an identical mass but a different radius would be a clear sign of hadron-quark phase transition in nuclear matter. Motivated by studies searching for significant deviations in the observables of twin stars, we investigate the differences that manifest in their $r$-mode instability windows and spin-down evolution. Firstly, we obtain a set of hybrid equations of state (which predict the existence of a third stable branch of compact objects) by employing the well-known Maxwell construction within the phenomenological framework of constant speed of sound parametrization. Then, we systematically study the influence of certain parameters, such as the energy density jump (in the resulting hybrid equation of state) and the crust elasticity, on the deviations appearing in the $r$-mode instability windows and spin-down evolution of twin stars. We conclude that two stars with an identical mass and fairly similar spin frequency and temperature, may behave differently with respect to $r$-modes. Thus, the future possible detection of gravitational waves (due to unstable $r$-modes) from a star laying in the stable region of the frequency-temperature plane would be a strong indication for the existence of twin stars. Furthermore, we consider current data for the spin frequencies and temperatures of observed pulsars and compare them to the predictions made from equations of state employed in this study. We find that, depending on the transition density and the rigidness of the crust, hybrid equations of state may be a viable solution for the explanation of existing data.


\end{abstract}

\maketitle

\section{Introduction}

Compact stars serve as excellent astrophysical laboratories for the study of dense nuclear matter \cite{Haensel-2007,Zeldovich-71,Weinberg-72,Schutz-85,Bielich-2020}. The systematic study of pulsars and the detection of gravitational waves (GW) have already yielded significant constraints on the nuclear equation of state (EOS) \cite{Demorest-2010,Fonceca-2016,Arzoumanian-2018,Antoniadis-2013,Cromartie-2020,Linares-2018,Abbott-2017,Abbott-2018,Abbott-2019}. A question that still remains unanswered concerns the relevant degrees of freedom up to densities appearing in neutron star cores \cite{Heiselberg-2000,Weber-2005}. Compact stars could be purely hadronic, but the very dense environment indicates the possible existence of exotic forms of matter such as deconfined quarks. The latter opens up new scenarios that predict strange quark stars, composed purely of strange quark matter, or hybrid stars where a quark core is surrounded by a mantle of hadronic matter. In practice, the distinction between neutron, strange and hybrid stars is not an easy task as their radius around the observed mass region of 1.4 $M_\odot$ is rather similar. Alternative approaches that may assist identifying the phases of nuclear matter within compact stars include the study of their thermal evolution \cite{Carvalho-2016,Lyra-2023}, binary neutron star mergers \cite{Prakash-2021,Bauswein-2020,Most-2020} and phenomena related to vibration or rotation \cite{Madsen-1992,Madsen-1998,Madsen-2000,Steiner-2008,Jaikumar-2008,Alford-2010,Ranea-2018,Bratton-2022}.

The construction of hybrid EOSs often requires us to describe the hadronic and quark phases separately. Depending on the dynamics of the phase transition, and mainly on the speed of sound structure in quark matter, a third family of compact objects may appear in the mass-radius plane~\cite{Bielich-2020}. The aforementioned family of compact stars gives rise to the existence of twin stars, i.e. stars with an identical mass but a fairly different radius \cite{Gerlach-1968,Kampfer-1981a,Kampfer-1981b,Glendenning-2000,Schertler-2000}. Recently, the scenario of twin stars drew a lot attention, mainly because of the discovery of GW and thus the possibility of detecting them \cite{Blaschke-2013,Alvarez-Castillo,Montana-2019,Tsaloukidis-2023,Alford-2013}. Note that identifying twins would be the smoking gun evidence of hadron-quark phase transition in compact stars. In a recent study, Lyra {\it et al.}~\cite{Lyra-2023} investigated the impact of compactness on the cooling of twin pairs, finding that only stars with significantly different radius exhibit considerable deviations in their thermal evolution. In addition, Tan {\it et al.}~\cite{Tan-2022a} examined imprints that manifest in binary Love universal relations due to the existence of a third family. Furthermore, Landry and Chakravarti \cite{Landry-2022} argued on the possibility of distinguishing twins, through their tidal deformabilities, with next-generation GW detectors. In the present work we study for the first time the deviations in the $r$-mode instability windows of twin stars and hence the differences that appear in their rotational limits.

It is well-established that relativistic stars may suffer a number of different  instabilities \cite{Kokkotas-2003a,Andersson-2003}. Among them, the $r$-mode instability (rotational mode) has been proposed as an explanation for the fact that neutron stars do not spin up to the theoretically allowed limit known as the Kepler frequency~\cite{Andersson-1998,Friedman-1998,Friedman-1999,Andersson-2001,Andersson-2003,Kokkotas-2003,Andersson-2001-b,Lindblom-2001}. The $r$-modes are oscillations appearing in rotating stars, and their restoring force is the Coriolis force. In principle, the $r$-mode instability can only take place if the gravitational-radiation driving timescale is shorter compared to the timescales of the various dissipation mechanisms that may occur in the neutron star interior. By equating the driving and damping timescales one obtains the so-called $r$-mode instability window, which defines a critical frequency (maximum spin frequency for stable $r$-modes) as a function of temperature~\cite{Andersson-1998}. 

In the past decades there has been an extensive study of the $r$-modes (and numerous other types of oscillation) due to the possible detection of their GW \cite{Andersson-1998,Friedman-1998,Friedman-1999,Andersson-2001,Andersson-2003,Kokkotas-2003,Andersson-2001-b,Lindblom-2001,Lindblom-98,Lindblom-1999a,Lindblom-2000a,Owen-98,Vidana-2012,Moustakidis-2015,Jyothilakshmi-2022}. There are several studies predicting that accreting stars in low mass x-ray binaries (LMXBs) may be subject to long-lasting $r$-modes~\cite{Anderson-1999,Reisenegger-2003}. In particular, compact stars containing exotic matter, such as deconfined quarks or hyperons, may be persistent sources of GW emission \cite{Anderson-2002,Nayyar-2006}. In addition, some authors~\cite{Chugunov-2014} argue for the existence of a large unobserved population of quiescent (postaccretion) LMXBs characterized by long-lived ($\sim$10$^9$ yr) $r$-mode emission. Specifically, Chugunov {\it et al.}~\cite{Chugunov-2014} suggested the existence of a new class of neutron stars, the so-called HOFNARS (hot and fast non accreting rotators). Such stars retain a high temperature due to heating associated with unstable $r$-modes. Following the discovery of gravitational radiation from binary neutron star mergers, the search for GW signals associated with $r$-modes has started~\cite{Caride-2019,Rajbhandari-2021,Covas-2022}. It is notable that, the absence of a detection so far has provided the opportunity to set upper limits on the GW emission and the $r$-mode saturation amplitude~\cite{Rajbhandari-2021,Covas-2022}. 

It has been shown that, the $r$-mode instability window of purely neutron stars is very wide to be compatible with current LMXBs data (assuming that all observed stars are stable with respect to $r$-modes, e.g. there are no HOFNARS) \cite{Zhou-2021,Haskell-2012}. Specifically, a very strong dissipation mechanism, such as a perfectly rigid crust, is essential for the stabilization of $r$-modes. Numerous studies have attempted to treat this problem by considering the presence of exotic degrees of freedom in compact star cores~\cite{Haskell-2012,Ofengeim-2019a,Ofengeim-2019b,Alford-2014,Pan-2006,Zheng-2006}. In particular, it has been shown that the bulk viscosity of hyperon or deconfined quark matter may be sufficient to stabilize $r$-modes for the frequencies and temperatures of the observed pulsars~\cite{Haskell-2012,Ofengeim-2019a}. However, it is important to comment that hyperons are expected to appear in densities of $2-3n_0$ (where $n_0=$ 0.16 fm$^{-3}$, is the nuclear saturation density). Thus, the fraction of the core where hyperons are present, and hence the effective damping due to their viscosity, is limited in low mass neutron stars. Similarly, the width of the $r$-mode instability window of hybrid stars is determined mainly by the amount of quark matter in their core~\cite{Jaikumar-2008,Alford-2012}. Subsequently, the fastest rotating pulsars can only be explained if they are massive enough~\cite{Ofengeim-2019a,Ofengeim-2019b}.

In Ref.~\cite{Papazoglou-2016}, the authors employed a set of analytical solutions of the Tolman-Oppenheimer-Volkov (TOV) equations in order to study the influence of neutron star bulk properties on the $r$-modes. They found that the instability window is quite sensitive to the radius of a star~\cite{Papazoglou-2016}. The latter leads to the conclusion that if twin stars do exist, their instability windows would deviate due to their radius difference. In addition, taking into account that the relevant degrees of freedom are different in the center of the two twins, the damping mechanisms that suppress the growth of the $r$-mode instability (bulk and shear viscosities) are going to be different as well~\cite{Madsen-1992}. This opens up a new intriguing scenario where two stars with an identical mass, and similar rotational frequency and temperature profiles, may behave differently with respect to $r$-modes. In particular, if we assume that a star having angular velocity $\Omega_i$ and temperature $T_i$ is stable with respect to the $r$-modes, then any other (same mass) star with similar temperature and $\Omega\leq\Omega_i$ should be stable as well. However, this is not necessarily the case if the two stars are twins since their instability windows are expected to be different. Thus, the future detection of GW due to unstable $r$-modes, from multiple sources, may allow us to identify a third family of compact objects.

The motivation of the present study is twofold. Firstly, we wish to systematically study the parameters (energy density gap, crust elasticity, transition density) that affect the deviation between the $r$-mode instability windows of twin stars. In addition, we wish to clarify how these parameters affect the differences that appear in the spin-down evolution (due to unstable $r$-modes) of twins. Secondly, we wish to examine if EOSs that predict a third family of compact objects are a viable solution for the explanation of current LMXBs data.

This paper is organized as it follows. Section~\ref{2} is devoted to the presentation of the hadronic models employed in this work and the construction of hybrid EOSs that predict twin star configurations. In Sec.~\ref{3} we discuss in detail the $r$-mode instability formalism, while in Sec.~\ref{4} we present a simplified model for the spin-down of compact stars (due to unstable $r$-modes). In Sec.~\ref{5} we present our results and discuss their implications. Sec.~\ref{6} contains a summary of our findings.

\section{Hadron-quark phase transition} \label{2}

A hybrid EOS often results from the combination of a low density hadronic model and a high density quark EOS. The key ingredient for the construction is the matching process between the two phases. In particular, there are two widely employed methods in order to obtain hybrid EOSs: a) the Maxwell construction and b) the Gibbs construction~\cite{Baym-2018}. The main difference of the aforementioned approaches is the number of charges that are globally conserved in the system~\cite{Glendenning-1992}.~In the former case, the phase transition is abrupt (i.e.~the two phases are separate), while in the latter scenario a mixed phase is present.

In the present work, we adopt the Maxwell construction for the description of the phase transition. According to lattice QCD calculations, this particular approach is the favored one in the scenario where the surface tension $\sigma_s$ in the hadron-quark crossover is larger than the critical value of $\mathrm{\sim 40 \;MeV\;fm^{-2}}$ and lower than the highest allowed one of $\mathrm{\sim 100 \; MeV\;fm^{-2}}$~\cite{Mariani-2017}. In this case the phase transition is sharp, resulting in a discontinuity in the energy density. Specifically, the energy density reads~\cite{Blaschke-2013,Alvarez-Castillo,Montana-2019,Alford-2013}
\begin{equation}
  \mathcal{E}(P) = \begin{cases} 
      \mathcal{E}_{\rm HADRON}(P), & P\leq P_{\rm tr} \\
      \mathcal{E}(P_{{\rm tr}}) + \Delta \mathcal{E} + (c_s/c)^{-2}(P-P_{{\rm tr}}), & P > P_{{\rm tr}}.
   \end{cases}
   \label{1}
\end{equation}
 where $P$ stands for the pressure, $c_s$ is the speed of sound and $c$ is the speed of light. Furthermore, $P_{tr}$ and $\Delta\mathcal{E}$ denote the transition pressure and the energy density jump, respectively. It is important to comment that the first line of Eq.~(\ref{1}) refers to the hadronic phase, while the second one to the quark model. We treat the quark phase using a phenomenological approach known as the constant speed of sound (CSS) parametrization~\cite{Alford-2014b}. More precisely, the second line of Eq.~(\ref{1}) can be though as a first order Taylor expansion of the energy density around the transition pressure. Even though such a treatment lacks a rigorous theoretical basis, it is widely employed as it is mimics the dynamics of the phase transition and it also allows an easy construction of EOSs predicting twin star configurations~\cite{Alford-2014b,Christian-2019,Christian-2021,Christian-2022,Han-2019a,Li-2021,Sharifi-2021,Paschalidis-2018,Alford-2017,Deloudis-2021,Han-2020}. 
 
Following the assumption of recent works~\cite{Christian-2019,Christian-2021,Christian-2022,Han-2019a,Li-2021,Sharifi-2021,Paschalidis-2018,Alford-2017,Deloudis-2021,Han-2020}, the speed of sound is set equal to the speed of light in order to obtain EOSs consistent with the 2~$M_\odot$ constraint. However, it is important to comment that, according to perturbative QCD (pQCD) calculations, $c_s$ in quark matter tends to $c/\sqrt{3}$ from below at large densities (conformal limit)~\cite{Kurkela-2010,Hebeler-2013,Kurkela-2014}. Nevertheless, the applicability of the conformal limit is reliable for the density range beyond 40$n_0$, which is considered to be well-below the central density of compact stars~\cite{Tan-2022b}. Therefore, the speed of sound in strange quark matter could possibly be larger than $c/\sqrt{3}$ and then decrease, with increasing baryon density, in order to satisfy the constraints from pQCD. For a detailed picture concerning the speed of sound structure of hybrid EOSs the reader is referred to Refs.~\cite{Tan-2022b,Pal-2023}.

A first order phase transition, between hadronic and quark matter, is not sufficient by itself for the appearance of a third family of compact objects. In particular, the appearance of twin stars requires the existence of an unstable region in the $M$-$R$ diagram, where the mass decreases with increasing central pressure. The condition that needs to be satisfied in order to obtain a third family was first studied by Seidov \cite{Seidov-1971} and it is formulated as follows 
\begin{equation}
    3 P_{\rm tr }+3\mathcal{E}_1-2\mathcal{E}_2<0, 
    \label{eq2}
\end{equation}
where $\mathcal{E}_1\equiv \mathcal{E}(P_{{\rm tr}})$ and $\mathcal{E}_2\equiv \mathcal{E}(P_{{\rm tr}}) + \Delta \mathcal{E}$. Thus, by reorganizing Eq.~(\ref{eq2}) one obtains the minimum energy density jump for the existence of twin star configurations, which is written as
 \begin{equation}
\Delta \mathcal{E}_{\rm cr}=\frac{1}{2}\mathcal{E}_{\rm tr}+\frac{3}{2}P_{\rm tr}.
\label{eq:seidov_limit}
\end{equation}
For EOSs that predict $\Delta \mathcal{E}\geq\Delta \mathcal{E}_{\rm cr}$ two distinct stable branches may appear on the $M$-$R$ plane.

The resulting hybrid EOSs ought to be consistent with neutron star observations.~For example, if one assumes that the $\sim$ 1.4 $M_\odot$ compact stars involved in GW170817~\cite{Abbott-2017} or in PSR J0030+0451~\cite{Riley-2019} are purely hadronic, then the low density sector of the EOS has to satisfy tight constraints ($\Lambda_{1.4}=190^{+390}_{-120}$ and $R_{1.4}\leq 14$ km, where $\Lambda$ denotes the dimensionless tidal deformability)~\cite{Landry-2022}. On the other hand, if these compact objects are hybrid stars, the aforementioned constraints are lifted from the hadronic part of the EOS. In the present work we adopt the GRDF-DD2 (simply DD2 from now on for practical purposes)~\cite{Typel-2018} and the NL3~\cite{Lalazissis-1997} EOSs for the description of the low density phase. It is worth commenting that, both of these EOSs have been previously employed in the study of twin stars~\cite{Tsaloukidis-2023,Sen-2022}. Finally, for the description of the outer crust (in the case of the NL3 model) the well-known EOS of Baym {\it et al.}~\cite{Baym-1971} is employed.

\section{$R$-mode instability formalism} \label{3}
Hydrodynamics and the influence of  various dissipative processes define the time evolution  of the  {\it r}-modes according to the law  $ e^{i \omega t-t/\tau}$, where $\omega$ is the real part of the frequency, given by~\citep{Lindblom-98}
\begin{equation}
\omega=-\frac{(l-1)(l+2)}{l+1}\Omega. \label{omega-1}
\end{equation}
In Eq.~(\ref{omega-1}), $\Omega$ is the angular velocity of the unperturbed star
and $l$ defines the kind of mode~\citep{Lindblom-98}. In  the present study, we will consider the case $l=2$. 
The imaginary part $1/\tau$ is related to the effects of
gravitational radiation and the various kinds of  viscosity (shear, bulk, etc.) 
\cite{Lindblom-1999a,Lindblom-2000a,Owen-98,Lindblom-98}. We consider the case of  small-amplitude
limit where  a mode is a driven, damped harmonic oscillator and the 
exponential damping timescale  is given by
\begin{eqnarray}
\frac{1}{\tau(\Omega,T)}&=&\frac{1}{\tau_{_{GR}}(\Omega)}+
\frac{1}{\tau_{_{EL}}(\Omega,T)}+\frac{1}{\tau_{_{BV}}(\Omega,T)}\nonumber \\
&+&\frac{1}{\tau_{_{SV}}(\Omega,T)},
\label{t-1}
\end{eqnarray}
where $\tau_{_{GR}}$ is the gravitational radiation timescale, $\tau_{_{EL}}$ is the damping timescale due to viscous dissipation at the boundary layer of the rigid crust and fluid core and $\tau_{_{BV}}$,$\tau_{_{SV}}$ are the bulk and shear viscosity dissipation timescales respectively~\cite{Lindblom-1999a,Lindblom-2000a,Owen-98,Lindblom-98}. It is notable that there is a battle between the gravitational radiation, which tends to drive the $r$-mode unstable, and the various  dissipation mechanisms that induce stabilization.  
The  critical angular velocity $\Omega_{\rm c}$ (or critical spin frequency $f_{\rm c}=\Omega_{\rm c}/2\pi$), corresponds to the velocity  at which the two mechanisms (amplification and damping) are balanced and it is found through the equation $1/\tau(\Omega_c)=0$~\cite{Lindblom-1999a,Lindblom-2000a,Owen-98,Lindblom-98}. 

The contribution of gravitational radiation to the imaginary part of the frequency is given by the following expression
\cite{Lindblom-2000a,Lindblom-98}
\begin{eqnarray}
\frac{1}{\tau_{_{GR}}}&=&-\frac{32\pi G
\Omega^{2l+2}}{c^{2l+3}}\frac{(l-1)^{2l}}{[(2l+1)!!]^2}\left(\frac{l+2}{l+1}\right)^{2l+2}\nonumber \\
&\times&
\int_0^{R}\rho(r) r^{2l+2} dr \quad \left({\rm s}^{-1}\right), \label{tgr-1}
\end{eqnarray}
where $\rho(r)$ is the mass density profile of a star.

The bulk viscosity $\xi_{_{BV}}$ is the dominant damping mechanism at high temperatures~\cite{Lindblom-98}. It originates from the variations of pressure and density due to the pulsation modes and in nucleonic matter it is given by the formula~\cite{Lindblom-98} 
\begin{eqnarray}
\xi_{_{BV}}^H&=&6.0\times 10^{-59}\left(\frac{l+1}{2}\right)^2 \left(\frac{{\rm Hz}}{\Omega}  \right)^2  \nonumber\\
&\times&\left(\frac{\rho}{{\rm gr\ cm^{-3}}}\right)^{2} \left(\frac{T}{{\rm K}}\right)^{6} \quad ({\rm  gr \ cm^{-1}\  s^{-1}}).
 \label{bulk-2}
\end{eqnarray}
For quark matter, the bulk viscosity is mainly determined by the weak process $d+s\leftrightarrow{}u+s$ \cite{Madsen-1992}. Following the discussion of Refs.~\cite{Madsen-1992,Jaikumar-2008}, we will use an approximate expression which is appropriate for small  oscillations of the fluid and when $2\pi T\gg\delta\mu=\mu_s-\mu_d$. Specifically,
\begin{equation}
\xi_{_{BV}}^Q = \frac{\alpha T^2}{(\kappa\Omega)^2+\beta T^4} \quad  { (\rm g \  cm^{-1} \ s^{-1}),}
\label{bulk-quark}
\end{equation}
where
\[ \alpha T^2=6.66 \times 10^{20} \left( \frac{\mu_d}{\rm MeV}\right)^3\left( \frac{m_s}{\rm MeV}\right)^4 T_9^2 \quad  { (\rm g\ cm^{-1}\ s^{-3}),}  \]
\[ \beta T^4=3.57 \times 10^{-8} \left( \frac{\mu_d}{\rm MeV}\right)^6\left(1+ \frac{m_s^2}{4 \mu_d^2}\right)^2 T_9^4 \quad  { (\rm s^{-2}),}  \]
where $T_9=T/(10^9 {\rm K})$, $\mu_d$ and $\mu_s$ are the chemical potential of the down and strange quarks respectively, $m_s=$~100~MeV is the mass of the strange quark and $ \kappa=$ 2/3. Since our model for quark matter does not provide information about the chemical potential profiles we will rely on the approximate expression $\mu_d = 235$~MeV~$(\rho/\rho_0)^{1/3}$~\cite{Madsen-1992}, which has been employed in numerous $r$-mode studies~\cite{Madsen-1992,Madsen-2000,Anderson-2002,Pan-2006,Lindblom-1999a}. In the previous formula $\rho_0$ denotes the nuclear density and it is equal to $2.8\times10^{14}$ g cm$^{-3}$. Finally, the bulk viscosity timescale is given by~\cite{Lindblom-98,Vidana-2012}
\begin{eqnarray}
\frac{1}{\tau_{_{BV}}}&=&\frac{4\pi}{690}\left(\frac{\Omega}{\Omega_0} \right)^4R^{2l-2}\left(\int_0^R \rho(r) r^{2l+2} dr \right)^{-1}\nonumber\\
&\times&
\int_0^R \xi_{_{BV}} \left(\frac{r}{R}  \right)^6\left[ 1+0.86\left(\frac{r}{R}  \right)^2
\right]r^2 dr, \label{bulk-1}
\end{eqnarray}
where $\Omega_0=\sqrt{\pi G\overline{\rho}}$ and
$\overline{\rho}=3M/4\pi R^3$ is the mean density of the star.

The shear viscosity $\eta_{_{SV}}$ is the dominant mechanism at low temperature. This mechanism  is  due to  the momentum transport  when  particle-particle scattering processes take place. In particular, the viscosities associated with the neutron-neutron and electron-electron scattering are given respectively by \cite{Lindblom-2000a}
\begin{equation}
\eta_{nn}=347  \left(\frac{\rho}{{\rm gr\ cm^{-3}}}\right)^{9/4} \left(\frac{T}{{\rm K}}\right)^{-2}\quad({\rm g\ cm^{-1}\ s^{-1}}),
\label{eta-nn-1}
\end{equation}
\begin{equation}
\eta_{ee}=6.0\cdot 10^6 \left(\frac{\rho}{{\rm gr\ cm^{-3}}}\right)^2 \left(\frac{T}{{\rm K}}\right)^{-2}\quad({\rm g\ cm^{-1}\ s^{-1}}). \label{eta-ee-1}
\end{equation} 
For quark matter the shear viscosity is dominated by quark-quark scattering in QCD. Following Ref.~\cite{Jaikumar-2008} we have
\begin{equation}
\eta_{q}= 5\times 10^{15} \left(\frac{0.1}{\alpha_s}  \right)^{3/2}\left(\frac{\rho}{\rho_0}  \right)^{14/9}T_9^{-5/3} 
\quad  { (\rm g \  cm^{-1} \ s^{-1})},
\label{shera-quark}
\end{equation}
where $a_s$ is the coupling constant for the strong interaction. In the present work we will use a typical value of $a_s=$ 0.1.
The dissipation timescale  due to the shear viscosity is given by \citep{Lindblom-98}
\begin{eqnarray}
\frac{1}{\tau_{_{SV}}}&=&(l-1)(2l+1) \left(\int_0^R \rho(r) r^{2l+2} dr \right)^{-1}\nonumber\\
&\times&
\int_0^R \eta_{_{SV}} r^{2l} dr, \quad ({\rm s^{-1}}).
\label{shear-1}
\end{eqnarray}

In the special case where the dissipation effect due to the crust has been included, the corresponding timescale is given by~\cite{Lindblom-2000a}
\begin{eqnarray}
\tau_{_{EL}}&=&\frac{1}{2\Omega}\frac{2^{l+3/2}(l+1)!}{l(2l+1)!! {\cal
C}_l}\sqrt{\frac{2\Omega R_c^2\rho_{cr}}{\eta_{cr}}}\nonumber\\
&\times&
\int_0^{R_c}\frac{\rho(r)}{\rho_{cr}}\left(\frac{r}{R_c}\right)^{2l+2}
\frac{dr}{R_c} \quad ({\rm s}).
 \label{tv-1-mew}
\end{eqnarray}
In Eq.~(\ref{tv-1-mew}), $R_c$ is the core's radius, while $\eta_{cr}$ and $\rho_{cr}$ are the viscosity and density of the fluid at the crust-core interface respectively. The factor ${\cal C}_l$, for $l=2$,  takes the value ${\cal C}_2=0.080411$. The expression of Eq.~(\ref{tv-1-mew}) refers to  the case where the crust is rigid and consequently static in the rotating frame. However, in a more realistic scenario, the motion of the crust (due to the mechanical coupling with the core) induces an increase of the  timescale $\tau_{_{EL}}$  by a factor of $1/{\cal S}^2$, where ${\cal S}$ is the slippage factor defined as ${\cal S}=\Delta v/v$~\cite{Levin-2001}. In particular,  $v$ denotes the velocity of the core and  $\Delta v$ is the difference between the velocities in the inner edge of the crust and the outer edge of the core~\cite{Levin-2001}. 

Even though the $r$-mode instability manifests in rotating objects, the presented formalism treats stars as if they were spherically symmetric. In principle, the rotational effects on a compact star's shape are of~$\sim \Omega^2$ and they are typically ignored in most calculations of inertial modes (as in the present study). At the moment, there are not very detailed calculations, but in the few cases where the $r$-mode spectrum was studied in fast rotation there were no reports for any significant effect~\cite{Stergioulas-2021,Gaertig-2008,Kruger-2020a,Kruger-2020b}.
More specifically, earlier non-linear~\cite{Stergioulas-2021} and linear simulations~\cite{Gaertig-2008} for fast rotating stars did not report considerable variations. However, it should be pointed out that, these calculations were performed in the Cowling approximation~\cite{McDermott-1983}. More recent simulations, based on a fully general relativistic code~\cite{Kruger-2020a,Kruger-2020b}, did not demonstrate significant influence in the spectrum from the presence of higher order $\Omega$ terms both in the the equilibrium configuration and in the perturbation equations~\cite{Kokkotas}.

It has been shown that the critical frequency $\Omega_{\rm c}$ is quite sensitive to the radius of a star~\cite{Papazoglou-2016}. More precisely, it has been found that for relatively low and high values of  temperature, $\Omega_{\rm c}$ scales with the radius as   $\Omega_{\rm c}\sim R^{-3/2}$ and $\Omega_{\rm c}\sim R^{-3/4}$, respectively~\cite{Papazoglou-2016}. The latter leads to the conclusion that, the $r$-mode instability windows (the areas above $\Omega_c(T)$ curves) of twin stars are going to be different. Furthermore, if one takes into account that the damping mechanisms in quark matter are, in principal, stronger than those in hadronic matter (for moderate temperature values~\cite{Jaikumar-2008}), then the instability window of a hybrid star is expected to be shifted to larger $\Omega_{\rm c}$ compared to the one of its hadronic twin. Thus, there are two mechanisms which act additively and may drastically affect the instability windows of the two different branches. The above findings are essentially a strong motivation for investigating the possible identification of twin stars due to the implications of their different instability windows.

Another  crucial issue is the limitation of the instability window, at high frequencies, from the corresponding Kepler angular velocity  $\Omega_{\rm K}$ (the maximum rotation frequency of the star). Following Ref.~\cite{Lindblom-98}, the latter quantity is approximately given by $\Omega_{\rm K}=\frac{2}{3}\Omega_0$. It is worth mentioning that, several studies, which investigated possible  universal relations for the Kepler frequency, have verified the existence of a correlation between the angular velocity of a maximally rotating neutron star and the mean density of the corresponding (equal mass) spherical configuration $\overline{\rho}$~\cite{Friedman-1989,Haensel-2009,Haskell-2018,Koliogiannis-2020}. It is interesting that in the case of twin stars the Kepler frequency of the hybrid branch may be even 20~$\%$ higher compared to the hadronic branch, due to the different radius values.  This apparent differentiation at the upper limit of the instability window can by itself be a criterion for separating the two branches. Connecting the analysis presented above with the fact that newly born compact stars may rotate close to their mass shedding limit~\cite{Koliogiannis-2021}, we conclude that the spin-down evolution paths of twin stars are going to exhibit distinct deviations.

\section{Spin-down and cooling }\label{4}
We are now going to present a simplified model to describe the spin-down (due to unstable $r$-modes) of a hadronic or hybrid star simultaneously with its cooling. During the phase that the angular momentum is radiated away to infinity by gravitational waves, the angular velocity of a star evolves as follows~\cite{Owen-98}
\begin{equation}
\frac{d\Omega}{dt}=\frac{2\Omega}{\tau_{GR}}\frac{\alpha^2 Q}{1-\alpha^2Q},
\label{dOmegadt-1}
\end{equation}
where $\alpha$ is the dimensionless $r$-mode amplitude  parameter. This parameter strongly affects the $r$-mode evolution and usually takes values
in the large  interval  $\alpha=1-10^{-8}$. Moreover, $\alpha$ in general depends both on the viscosity (and consequently on the temperature $T$ and cooling process) and on time. However, following Refs.~\cite{Owen-98,Routray-2021} we consider that $d\alpha/dt=0$.  In addition, the quantity $Q$ is related to the bulk properties of a star and it is defined as $Q=3\tilde{J}/2\tilde{I}$ where
\begin{equation}
\tilde{J}=\frac{1}{MR^4}\int_{0}^R \rho(r)r^6 dr, \qquad \tilde{I}=\frac{8\pi}{2MR^2}\int_{0}^R \rho(r)r^4 dr.
\label{I-J-1}
\end{equation}
Under the aforementioned assumptions, on can solve Eq.~(\ref{dOmegadt-1}) analytically and obtain~\cite{Moustakidis-2015,Routray-2021}
\begin{equation}
\Omega(t)=\left(\frac{1}{\Omega^{-6}_{in} -6{\cal D}t}\right)^{1/6},
\label{dOmegadt-2}
\end{equation}
where
\begin{equation}
{\cal D}=\frac{2\alpha^2Q}{\tilde{\tau}_{GR} (1-\alpha^2Q)}\frac{1}{\Omega_0^6}, \quad \tilde{\tau}_{GR}=\left(\frac{\Omega}{\Omega_0}\right)^{6}\tau_{GR}.
\label{Cons-1}
\end{equation}
$\Omega_{in}$ is a free parameter that corresponds to the initial angular velocity and $\tilde{\tau}_{GR}$ is the fiducial gravitational-radiation timescale. 

In order to combine the concurrent processes  of spin-down and thermal evolution of twin stars, we adopt, as a first approximation, the standard cooling scenario~\cite{Page-2004}. Standard cooling is the simplest thermal evolution model, since it only accounts for the energy loss due to neutrino emission via the modified Urca process~\cite{Page-2004}. Such an approach has been previously employed by Owen {\it et al.}~\cite{Owen-98} and the temperature drops according to the law
\begin{equation}
T(t)=\left(\frac{t}{t_c}+\left(\frac{10^9 \ {\rm K}}{T_i}  \right)^6   \right)^{-1/6} 10^9 \ {\rm K},
 \label{T-t}   
\end{equation}
\begin{figure*}
  \includegraphics[width=\textwidth,scale=1]{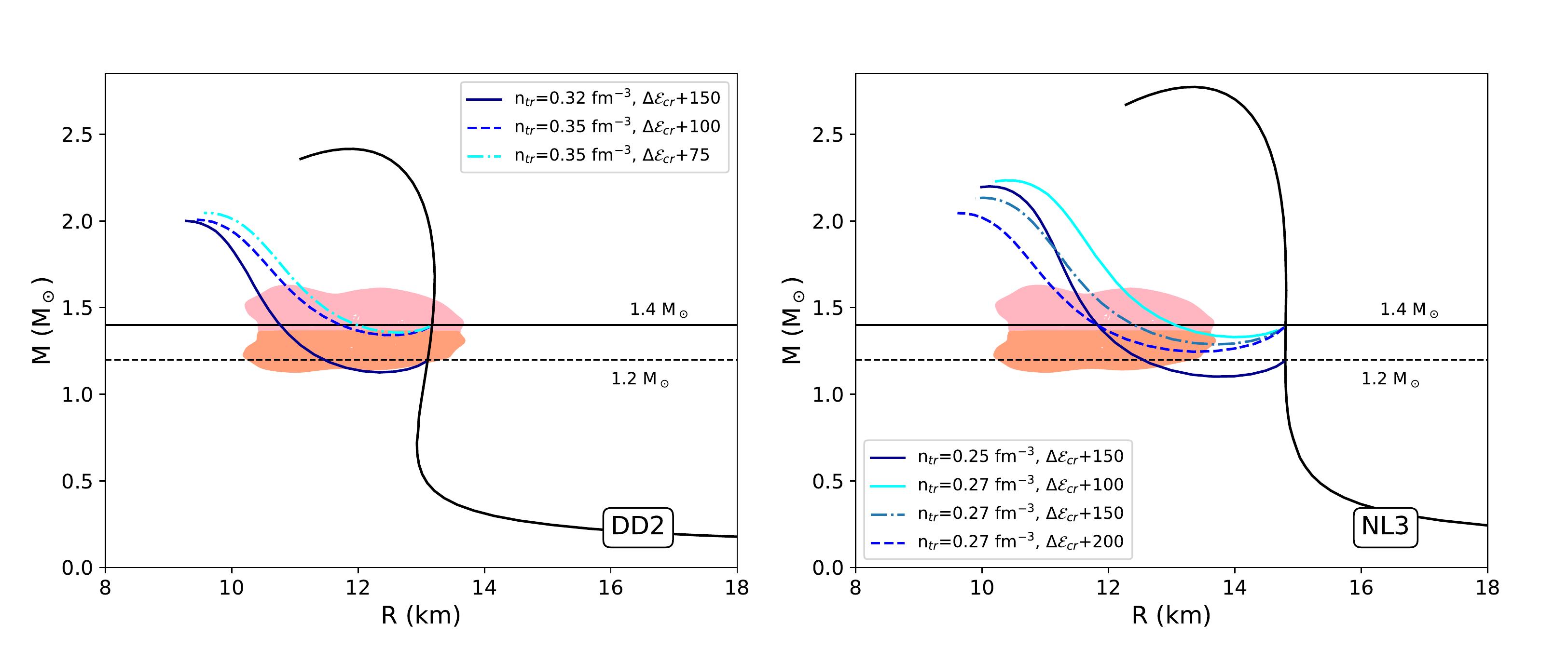}
  \caption{Mass-radius diagrams for the DD2 (left panel) and NL3 (right panel) EOSs. The black solid curves indicate the original EOSs. The solid (dashed) horizontal line is set to 1.4 (1.2)~$M_\odot$. The shaded areas correspond to the constraints from the analysis of the GW170817 event \cite{Abbott-2017,Abbott-2018}. Each hybrid EOS is identified from the baryon density $n_{tr}$  where the phase transition occurs and by the energy density jump. The energy density gap is given in units of MeV fm$^{-3}$ which are omitted in the legend for simplicity.} 
  \label{Fig1}
\end{figure*}
where  $T_i$ is the initial temperature of the star (a typical value is $T_i\simeq 10^{11} \ {\rm K}$) and $t_c$ is the cooling rate parameter ($t_c\simeq 1\ {\rm year}$~\cite{Owen-98}). We need to highlight that, the present model is more suited for the case of young isolated neutron stars, as only neutrino and no photon emission is considered, and it may fail to provide an accurate description for the cooling of LMXBs. In particular, the thermal evolution of LMXBs is a far more complicated problem, as there may be additional mechanisms in action, such as heating due to accretion. For more details concerning the use of the present model in the $r$-mode evolution, the reader is referred to the discussion of Ref.~\cite{Owen-98}.

One may argue that the two twins may cool down in a different way, considering the different cooling mechanisms in hadronic and quark matter. Obviously, a more elaborate study is necessary if one is interested in an accurate quantitative description of the cooling process. However, according to the findings of Lyra {\it et al.}~\cite{Lyra-2023}, the thermal evolution of twin stars is only distinct when there is a large difference in their compactness. More precisely, in the case where there is a 10~$\%$ compactness difference (which is the case for the configurations constructed in the present study), the thermal evolution of the two twins is nearly identical~\cite{Lyra-2023}. From that perspective, we expect that the selected model will allow a qualitative comparison for the evolution of twin stars on the $f-T^\infty$ plane.

\section{Results and discussion} \label{5}
\subsection{Mass-radius diagrams}
In order to study the differences that manifest in the $r$-mode instability windows and spin evolution of twin stars we constructed a set of hybrid EOSs, using the analysis presented in Sec.~\ref{2}. In particular, the low density phase is described by the DD2 and NL3 EOSs, where for the quark matter a phenomenological constant speed of sound model is employed. The values of the energy jump are selected in order to obtain EOSs that are consistent with the constraints from astrophysical observations. For both hadronic models, the resulting EOSs predict twin stars with mass of 1.2 or 1.4 $M_\odot$.

Figure~\ref{Fig1} depicts the mass-radius dependence for the EOSs employed in this study.~In the left panel the hadronic phase is described using the DD2 EOS, while for the results of the right panel the NL3 model was employed. The solid black curves stand for the case where no phase transition occurs (i.e. the $M$-$R$ diagrams for the purely hadronic EOSs).~In addition, the shaded areas correspond to constraints based on the analysis of the GW170817 event~\cite{Abbott-2017,Abbott-2018}. Finally, the horizontal lines are drawn to indicate the twin configurations with 1.2 and 1.4 $M_\odot$. 

As it is evident from Fig.~\ref{Fig1}, increasing the energy density jump results into a softening of the EOS. Thus, the largest values for $\Delta\mathcal{E}$ are selected so that EOSs remain consistent with the 2 $M_\odot$ constraint. Furthermore, we need to highlight that as $\Delta\mathcal{E}$ increases the radius difference of the two twins becomes larger. The latter is expected to play a critical role concerning the deviation of the $r$-mode instability windows (see Sec.~\ref{3})~\cite{Papazoglou-2016}. It is important to note that our analysis does not include the limiting case where $\Delta\mathcal{E}=\Delta\mathcal{E}_{cr}$, as in such a scenario the separation of the two twins is almost negligible. In particular, if the phase transition occurs in relatively low baryon density a third family may not even appear \cite{Tsaloukidis-2023}.

\subsection{Qualitative analysis}

\begin{figure}[t]
  \centering
  \includegraphics[width=9.5 cm,scale=1]{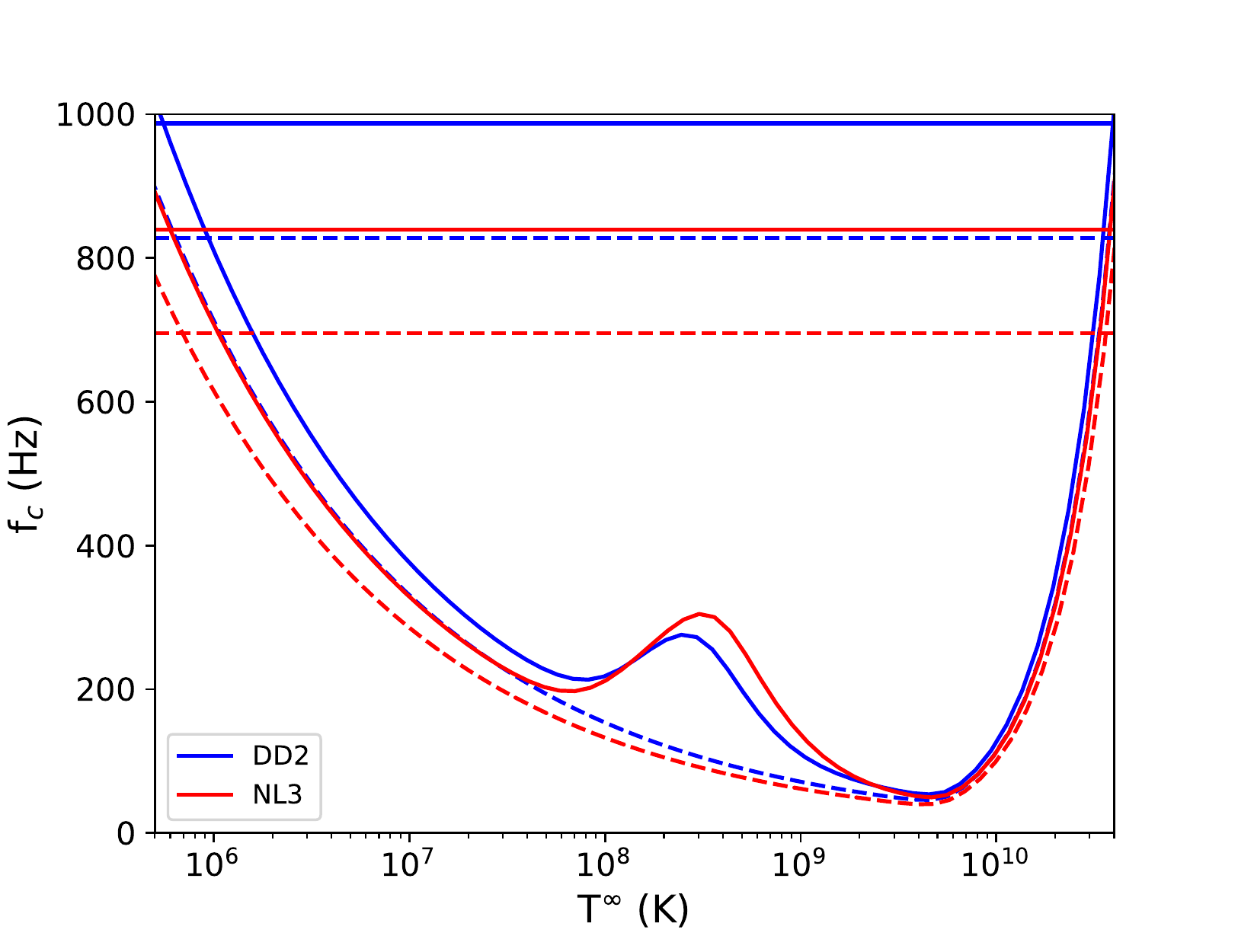}
  \caption{Critical spin frequency $f_c$ as a function of the redshifted temperature $T^\infty$ ($r$-mode instability windows) for 1.4 $M_\odot$ twin stars for the DD2 (blue) and NL3 (red) EOSs. The dashed lines and solid lines correspond to the hadronic and hybrid twins, respectively. The horizontal lines denoted the Kepler frequency for its star. The value for the energy density gap is $\Delta\mathcal{E}_{cr}+$100 MeV fm$^{-3}$ for both EOSs.}
  \label{Fig2}
\end{figure}

Figure~\ref{Fig2} presents the $r$-mode instability windows of 1.4~$M_\odot$ twin stars for the case where $\Delta\mathcal{E}=\Delta\mathcal{E}_{cr}+$100 MeV fm$^{-3}$. The damping mechanism due to the presence of a solid crust is not included. The results for the hadronic and hybrid stars are indicated using dashed and solid curves, respectively. Additionally, the horizontal lines stand for the corresponding Kepler frequencies. Note that in the x axis of the plot, one would not find the temperature $T$, appearing in the formalism of Sec.~\ref{3}, but the so-called redshifted temperature which is given by $T^\infty=T\sqrt{1-2C},$
where $C=GM/Rc^2$ is the compactness of a star. 

Firstly, we need to underline the sensitivity of the instability window to the employed EOS. Specifically, by comparing the instability windows of the purely hadronic configurations one finds that the predicted critical frequency is lower (in the low temperature region) when the NL3 EOS is employed. This results from the fact that the radius of a 1.4 $M_\odot$ compact star is smaller when the DD2 model is used~\cite{Papazoglou-2016}. Incidentally, the radius of the hybrid star constructed using the NL3 EOS coincides with the radius of the hadronic configuration using the DD2 EOS. The latter results into an overlap of their instability windows in the low temperature regime. However, the existence of a quark core, in the hybrid star, leads to significant differences in the critical frequencies for $T^\infty\geq$ 10$^8$ K, where the bulk viscosity plays a crucial role~\cite{Jaikumar-2008}. 

For a qualitative comparison of the $r$-mode instability windows of twin pairs one can divide Fig.~\ref{Fig2} into three representative regions. In particular, for $T^\infty\leq$~10$^8$~K (where the shear viscosity is the dominant dissipation mechanism~\cite{Jaikumar-2008}), the radius difference plays a crucial role for the apparent critical frequency deviations. For 10$^8$~K~$\leq T^\infty\leq$~10$^{10}$~K, the bulk viscosity (of quark matter~\cite{Jaikumar-2008}) is the major damping mechanism and the trend of the $f_c(T^\infty)$ curve is altered for the hybrid twin. More precisely, the critical frequency increases and then decreases with temperature leading to a local maxima. This topological difference derives from the fact that the bulk viscosity of quark matter is not a monotononic function of temperature. Finally, for $T^\infty\geq$ 10$^{10}$ K the bulk viscosity of hadronic matter dominates in both twins and their instability windows essentially coincide. From an observational perspective, the differences that appear in the low temperature regime will lead to different limits on the spin up of accreting pulsars in LMXBs, depending on whether they are hybrid or purely hadronic. Apart from the deviations appearing in the critical spin frequencies of twin stars, we need to comment that there is a~$\sim$~17~$\%$ difference in their Kepler velocities as well. Subsequently, a young hybrid star can rotate much faster than its hadronic twin.

\subsection{Energy density jump and crust effects}

At this point we wish to systematically study the influence of certain parameters on the instability window deviations of twin stars. In particular, we are going to vary the value of the energy density gap and examine its effects. Furthermore, up to this point, the only dissipative mechanisms considered in our calculations were the bulk and shear viscosities. Now, we are also going to include the damping mechanism due to the presence of a viscous boundary layer. It is interesting that, as the aforementioned mechanism is strong and common for both twins, the critical frequency deviations due to different viscosities are expected to be less pronounced.

\begin{figure}
  \centering
  \includegraphics[width=8 cm,scale=0.8]{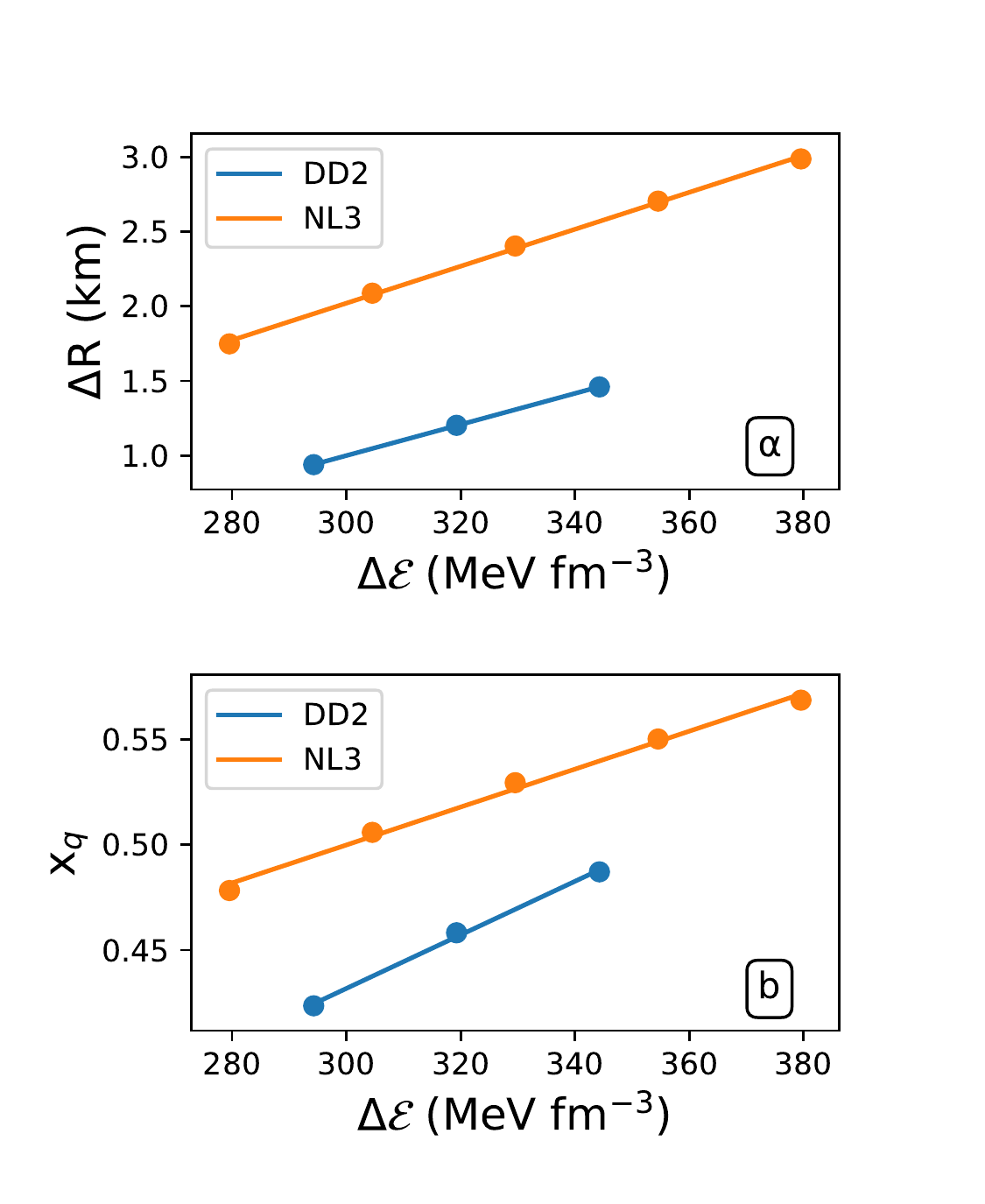}
  \caption{Panel a: Radius difference between 1.4 $M_\odot$ twin stars as a function of the energy density gap, Panel b: The quark core fraction of a 1.4 $M_\odot$ hybrid star as a function of the energy density jump. }
  \label{Fig3}
\end{figure}

Firstly, we are going to investigate the importance of the energy density jump. As we mentioned, $\Delta\mathcal{E}$ is the regulator of the radius difference between twin configurations. Figure~\ref{Fig3}(a) depicts the dependence of $\Delta R$ on $\Delta \mathcal{E}$ for 1.4 $M_\odot$ twin stars. Surprisingly, we find that the aforementioned quantities are connected through a linear formula. Even though the exact $\Delta R$-$\Delta \mathcal{E}$ relation is sensitive to the low density model, the slopes of the resulting fitted lines appear to be very similar. It is worth pointing out that, from the analysis presented in the previous sections, an increment of $\Delta \mathcal{E}$ will result into larger deviations in the instability windows due to an increase of $\Delta R$. However, as it is evident from Fig.~\ref{Fig3}(b), a larger value of $\Delta \mathcal{E}$ also results into a hybrid twin with a larger quark core fraction $x_q=R_q/R$ (where $R_q$ is the quark core radius). Hence, the damping due to quark matter's bulk viscosity is going to be even more effective. It is noteworthy that $\Delta \mathcal{E}$ and $x_q$ are also linearly dependent and that the slopes of the lines are, once again, not strongly sensitive to employed hadronic model. The relations presented in Fig.~\ref{Fig3} can be added to the other correlations found in the detailed analysis of Ref.~\cite{Sen-2022}. Finally, we need to underline that, through the relations found above, the knowledge of the radius difference of twin stars may provide important information concerning the phase transition and the interior of hybrid stars. 

\begin{figure*}
  \includegraphics[width=\textwidth,scale=0.8]{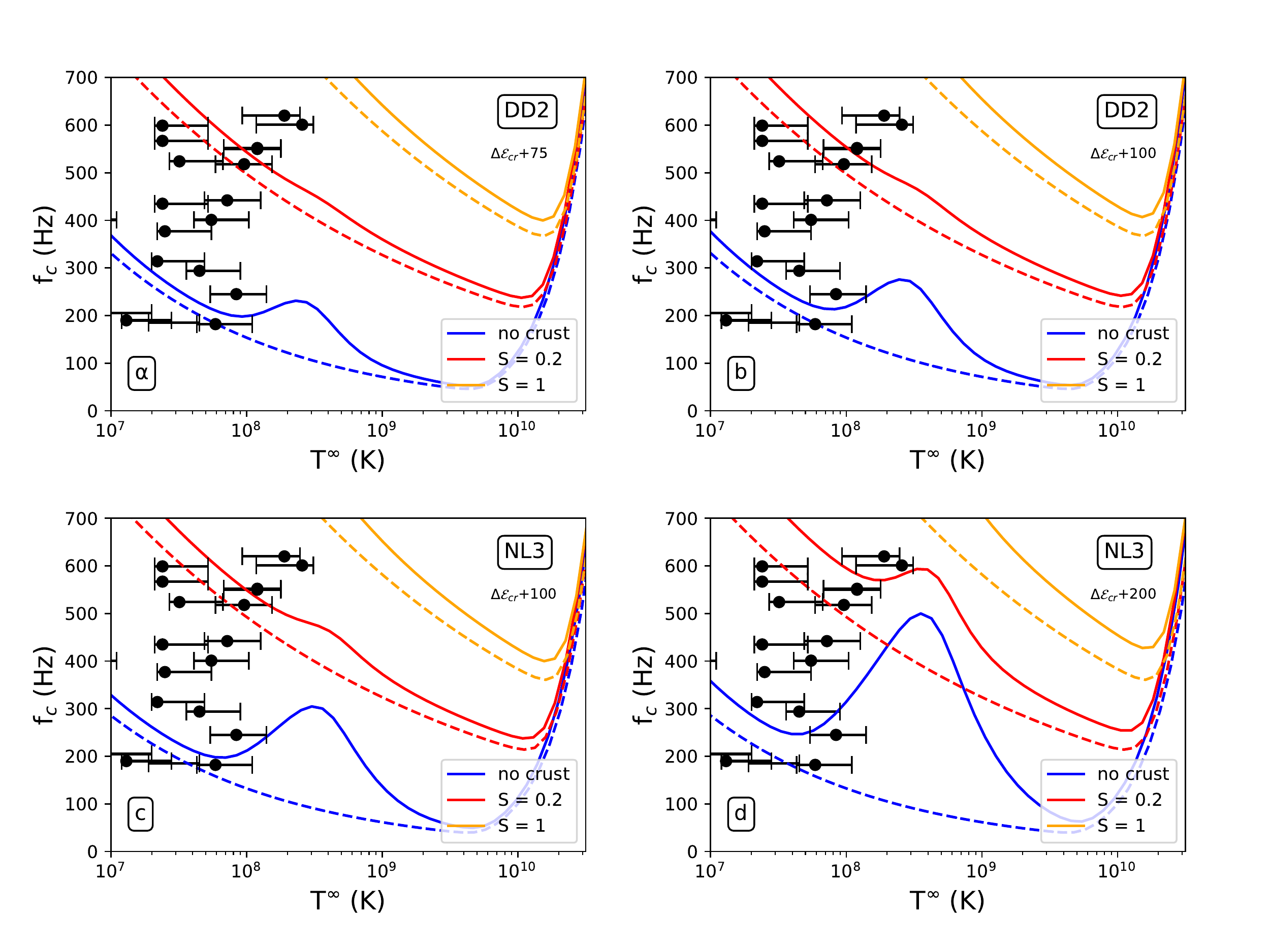}
  \caption{The effect of energy density gap in the deviation of the $r$-mode instability windows of 1.4~$M_\odot$ twin stars for increasing crust elasticity values. Panel (a) DD2 EOS and $\Delta\mathcal{E} = \Delta\mathcal{E}_{cr}+$75 MeV fm$^{-3}$, 
  (b) DD2 EOS and $\Delta\mathcal{E} = \Delta\mathcal{E}_{cr}+$100 MeV fm$^{-3}$, (c) NL3 EOS and $\Delta\mathcal{E} = \Delta\mathcal{E}_{cr}+$100 MeV fm$^{-3}$, (d) NL3 EOS and $\Delta\mathcal{E} = \Delta\mathcal{E}_{cr}+$100 MeV fm$^{-3}$. Dashed (solid) lines indicate the hadronic (hybrid) twin. The dotted points correspond to observational data taken from Ref.~\cite{Gusakov-2014}. The no crust indication in the legend corresponds to the case where the damping due to a viscous boundary layer has not been included.}
  \label{Fig5}
\end{figure*}

Figure~\ref{Fig5} depicts the dependence of the critical frequency on temperature, for 1.4 $M_\odot$ twin stars constructed using different $\Delta \mathcal{E}$ values. In addition, the crust elasticity $\mathcal{S}$ is varied from 0 to 1 in order to investigate the effects of a viscous boundary layer. Furthermore, Fig.~\ref{Fig5} contains observational data inferred from LMXBs and millisecond pulsars, where the temperature uncertainties derive from different assumptions concerning the star's envelope composition \cite{Gusakov-2014}. In particular, the small filled circles demonstrate the internal temperature when a partially accreted envelope is considered ($T_{fid}$ in Table 1 of Ref.~\cite{Gusakov-2014}), while the error bars stand for the cases of a fully accreted and a pure iron envelope ($T_{acc}$ and $T_{Fe}$ columns in Table 1 of Ref.~\cite{Gusakov-2014}).
\begin{table}[t]
\caption{The difference in the critical frequencies for 1.4 M$_\odot$ twin stars using the NL3 model for different values of temperature and energy density jump. The damping due to a viscous boundary layer (rigid crust) is not considered for the results presented in this table.}
\begin{center}
\setlength\tabcolsep{0pt}
\begin{tabular*}{\linewidth}{@{\extracolsep{\fill}} cccc }
    \hline
    \hline
     $\Delta\mathcal{E}$ (MeV fm$^{-3}$) & $\Delta R$ (km) & $T^\infty$ (10$^8$ K) & $\Delta f_c$ (Hz)\\
    \hline
    \multirow{3}{*}{ $\Delta\mathcal{E}_{cr}+$100} & & 1 & 78.98 \\ & 1.75 & 5 & 179.66\\ & & 10 & 75.14\\ 
    \hline
    \multirow{3}{*}{$\Delta\mathcal{E}_{cr}+$150} &  & 1 & 132.58 \\ & 2.40 & 5 & 288.95 \\ & & 10 & 137.42\\
    \hline
    \multirow{3}{*}{$\Delta\mathcal{E}_{cr}+$200} & & 1 & 181.99 \\ & 2.99 & 5 & 381.43\\ & & 10 & 192.05\\ 
    \hline
    \hline
\end{tabular*}
\end{center}
\label{t1}
\end{table}

\begin{table}
\caption{The difference in the critical frequencies for 1.4~$M_\odot$ twin stars for different values of temperature and crust elasticity. The results were obtained using the NL3 model with $\Delta\mathcal{E}=\Delta\mathcal{E}_{cr}+$200 MeV fm$^{-3}$. The results for
this EOS, in the case where the crust damping mechanism is
not considered, can be found in Table~\ref{t1}}
\begin{center}
\begin{tabular*}{\linewidth}{@{\extracolsep{\fill}} ccc }
    \hline
    \hline
     $S$  & $T^\infty$ (10$^8$ K) & $\Delta f_c$ (Hz)\\
    \hline
    \multirow{3}{*}{0.2}   & 1 & 102.279 \\  & 5 & 199.014\\  & 10 & 103.426\\
    \hline
    \multirow{3}{*}{1}  & 1 & 165.56 \\  & 5 & 169.475\\  & 10 & 129.472\\ 
    \hline
    \hline
\end{tabular*}
\end{center}
\label{t2}
\end{table}
It is worth pointing out that, the instability window differences are more pronounced in the case where the NL3 model is employed. This results from the fact that as the NL3 model is stiffer, it allows the construction of EOSs that satisfy observational constraints even for large $\Delta \mathcal{E}$ values. Furthermore, Fig.~\ref{Fig5} illustrates the strong impact of $\Delta \mathcal{E}$ on the resulting $r$-mode instability window of the hybrid twin. In particular, in the case of $\Delta \mathcal{E}_{cr}$ + 200, the spin frequency difference for the two twins may reach values of $\sim$ 400 Hz (see Fig.~\ref{Fig5}(d) and Table \ref{t1}). In accordance to the results present by Lyra {\it et al.} \cite{Lyra-2023}, we conclude that the role of the compactness is not only critical concerning the thermal evolution of twin pairs, but it also significantly affects the $r$-mode instability window of the hybrid configuration. 

The most important effect when the damping due to a solid crust is included, is that the peak appearing in the instability window of the hybrid star (in a temperature region around $\sim$ 3$\times$10$^8$ K) drops down. However, depending on the selected $\Delta \mathcal{E}$ value, large $f_c$ differences for the two twins may remain (see Fig.~\ref{Fig5} and Table~\ref{t2}).

\subsection{Comparison with observational data}

 \begin{figure*} 
  \includegraphics[width=\textwidth,scale=1]{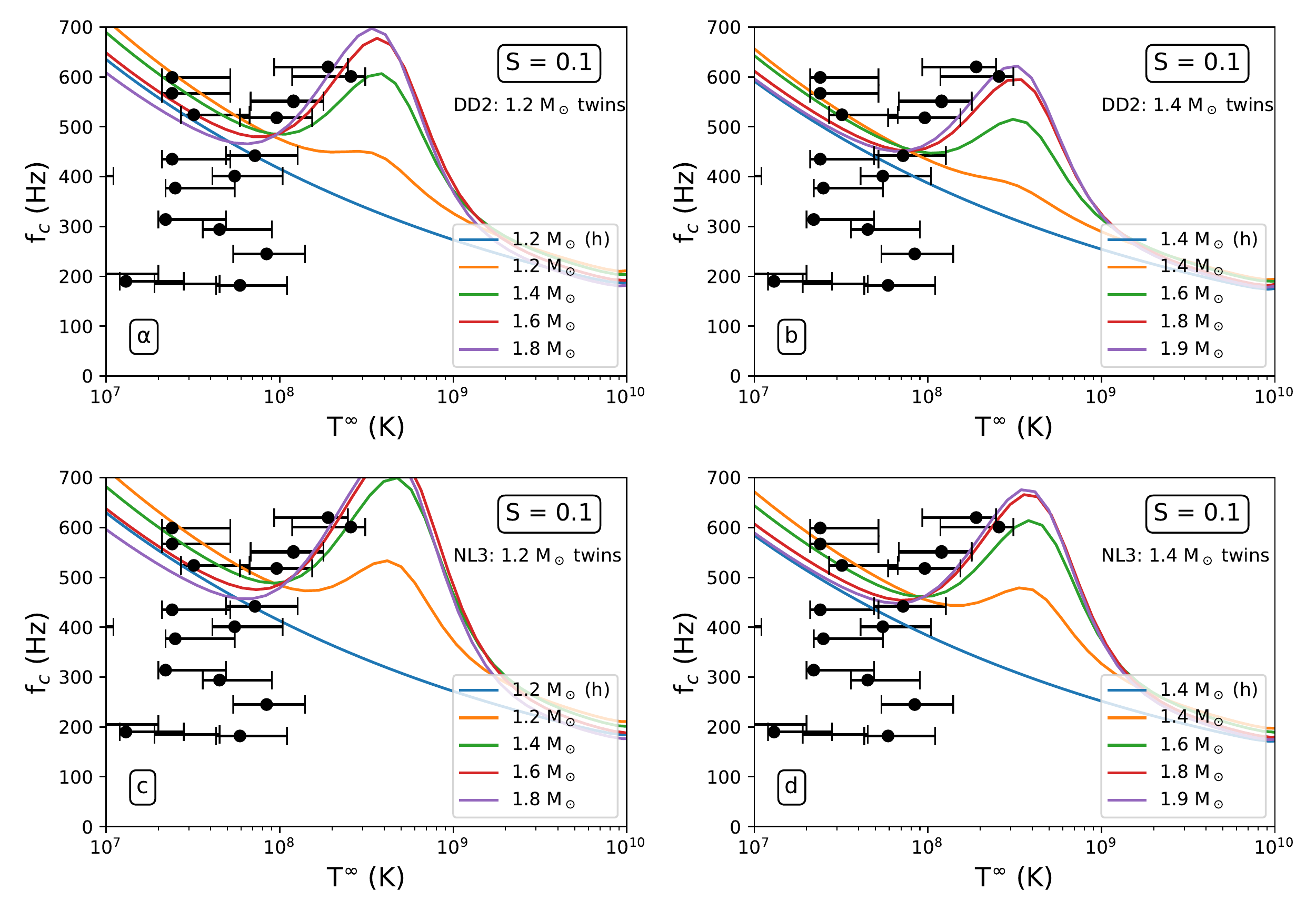}
  \caption{The $R$-mode instability windows of compact stars in the mass range 1.2$-$1.9 $M_\odot$. The EOSs used are: (a) DD2, $n_{tr}=$~0.32~fm$^{-3}$, $\Delta \mathcal{E}_{cr}+$ 150, (b) DD2, $n_{tr}=$ 0.35 fm$^{-3}$, $\Delta \mathcal{E}_{cr}+$ 100, (c) NL3, $n_{tr}=$ 0.25 fm$^{-3}$, $\Delta \mathcal{E}_{cr}+$ 150, (d) NL3, $n_{tr}=$ 0.27 fm$^{-3}$, $\Delta \mathcal{E}_{cr}+$ 150. The dotted points correspond to observational data, and they are taken from Ref.~\cite{Gusakov-2014}. The (h) appearing in the legend stands for the most massive purely hadronic configuration. The value for the crust elasticity is considered to be $\mathcal{S}=$~0.1.}
  \label{Fig6}
\end{figure*}
As previously mentioned, it is rather difficult to explain the observational data in the context of a purely hadronic star. In particular, the not realistic assumption of a perfectly rigid crust is essential \cite{Haskell-2012}. For that matter, several studies have investigated the $r$-mode instability window of compact stars containing exotic forms of matter~\cite{Haskell-2012,Ofengeim-2019a,Ofengeim-2019b,Alford-2014,Pan-2006,Zheng-2006}. In a recent work, Ofengeim {\it et al.} \cite{Ofengeim-2019a,Ofengeim-2019b} examined if the existence of hyperons in the core of compact stars can lead to results compatible with current LMXBs data. What they found is that for neutron stars with $M\leq 1.9$~$M_\odot$, the bulk viscosity of hyperonic matter leads to $r$-mode stabilization in the $f-T^\infty$ regime where the observed neutron stars appear \cite{Ofengeim-2019a,Ofengeim-2019b}.

At this point we wish to examine if the hybrid EOSs constructed in this study are in accordance to the observed spin frequencies and temperatures in LMXBs. As it is evident from Fig.~\ref{Fig5}, the $r$-mode instability window of the hybrid twin is always narrower. In addition, in a minimal scenario where the effects of the crust are not included the explanation of the observational data is not possible. However, depending on the energy density jump, a moderate crust elasticity value would suffice for the construction of instability windows compatible with observations. Specifically, for the NL3 model ($\Delta \mathcal{E}_{cr}+$ 200) and a relatively small crust elasticity of 0.2, most of the observed stars lay in the stable region of the $f-T^\infty$ plane. Another critical point is that, in all cases there are stars (from the dataset) that lay in the region between the $f_c(T^\infty)$ curves for the two twins. Hence, while such stars can be considered stable with respect to $r$-modes in the framework of the hybrid twin, they would be unstable if they were purely hadronic. The latter comment is of most importance, as the detection of GW emission, from stars laying in a $f-T^\infty$ region where $r$-modes are considered to be stable, would be a strong indication of hadron-quark phase transition.

In Fig.~\ref{Fig6} we present the $r$-mode instability windows for compact stars in the mass range 1.2$-$1.9 $M_\odot$ and a relatively low crust elasticity value $\mathcal{S}=$ 0.1. In panels a and c the twin configurations have a mass of 1.2 $M_\odot$, while in panels b and d their mass is 1.4 $M_\odot$. In the first case we find that, the bulk viscosity of quark matter is sufficient to stabilize $r$-modes for moderately massive compacts stars ($M \leq 1.6$~$ M_\odot$) in the whole $f-T^\infty$ range occupied by the observed stars in LMXBs. In latter case, where the phase transition occurs at higher baryon density, more massive compact star configurations (1.8 or 1.9 $M_\odot$ depending on the hadronic EOS) are essential for the explanation of current LMXBs data. 
 \begin{figure*}
  \includegraphics[width=\textwidth,scale=1]{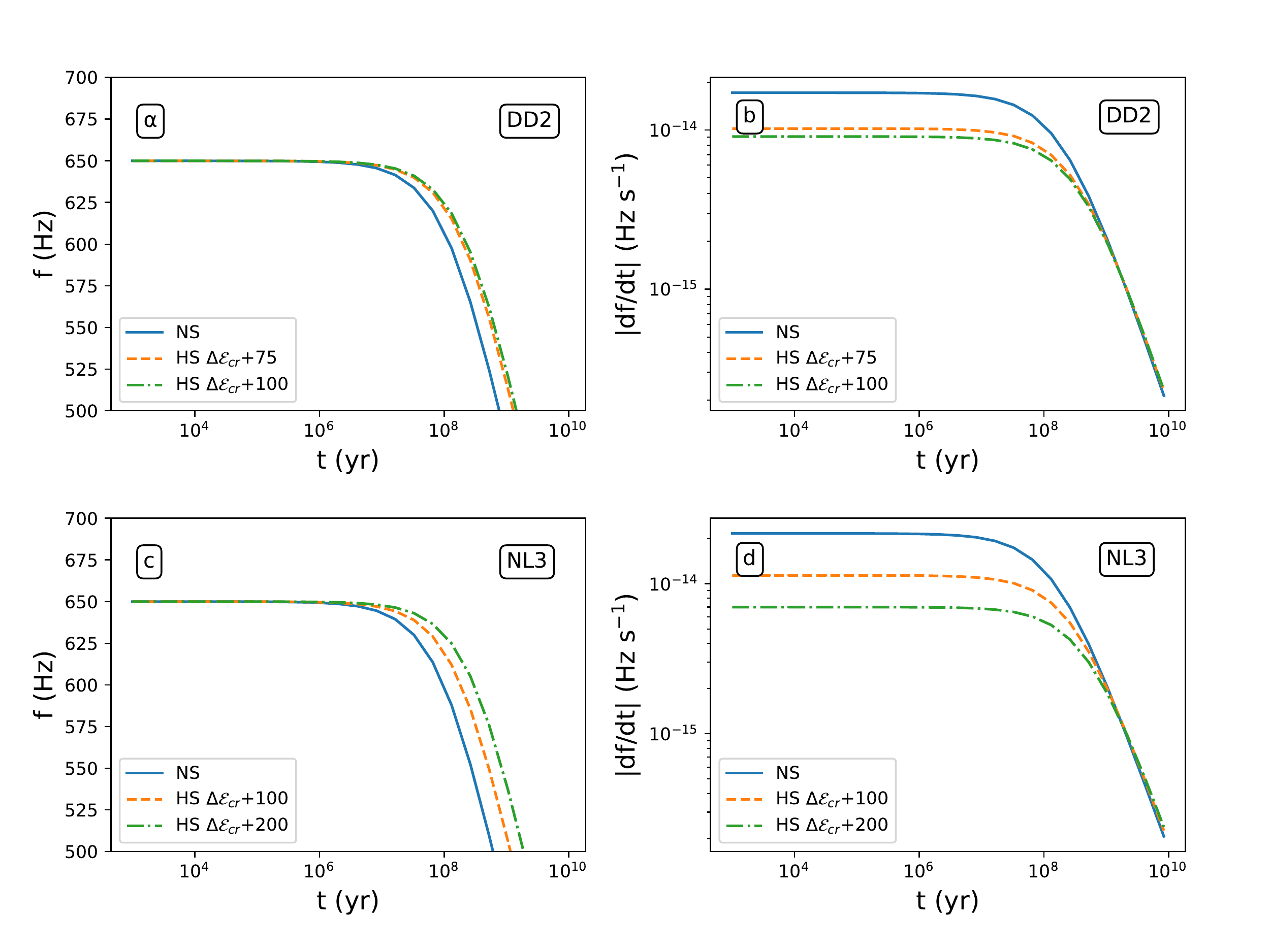}
  \caption{Panel: (a) Spin frequency as a function of time for 1.4 $M_\odot$ twin stars constructed with the DD2 EOS, (b) Spin-down rate as a function of time for 1.4 $M_\odot$ twin stars constructed with the DD2 EOS, (c) Spin frequency as a function of time for 1.4 $M_\odot$ twin stars constructed with the DD2 EOS, (d) Spin-down rate as a function of time for 1.4 $M_\odot$ twin stars constructed with the NL3 EOS. In all panels two different values for $\Delta\mathcal{E}$ were used (see legends).
}
  \label{spin-down-1}
\end{figure*}
 \begin{figure}
  \centering
  \includegraphics[width=9.5 cm ,scale=1]{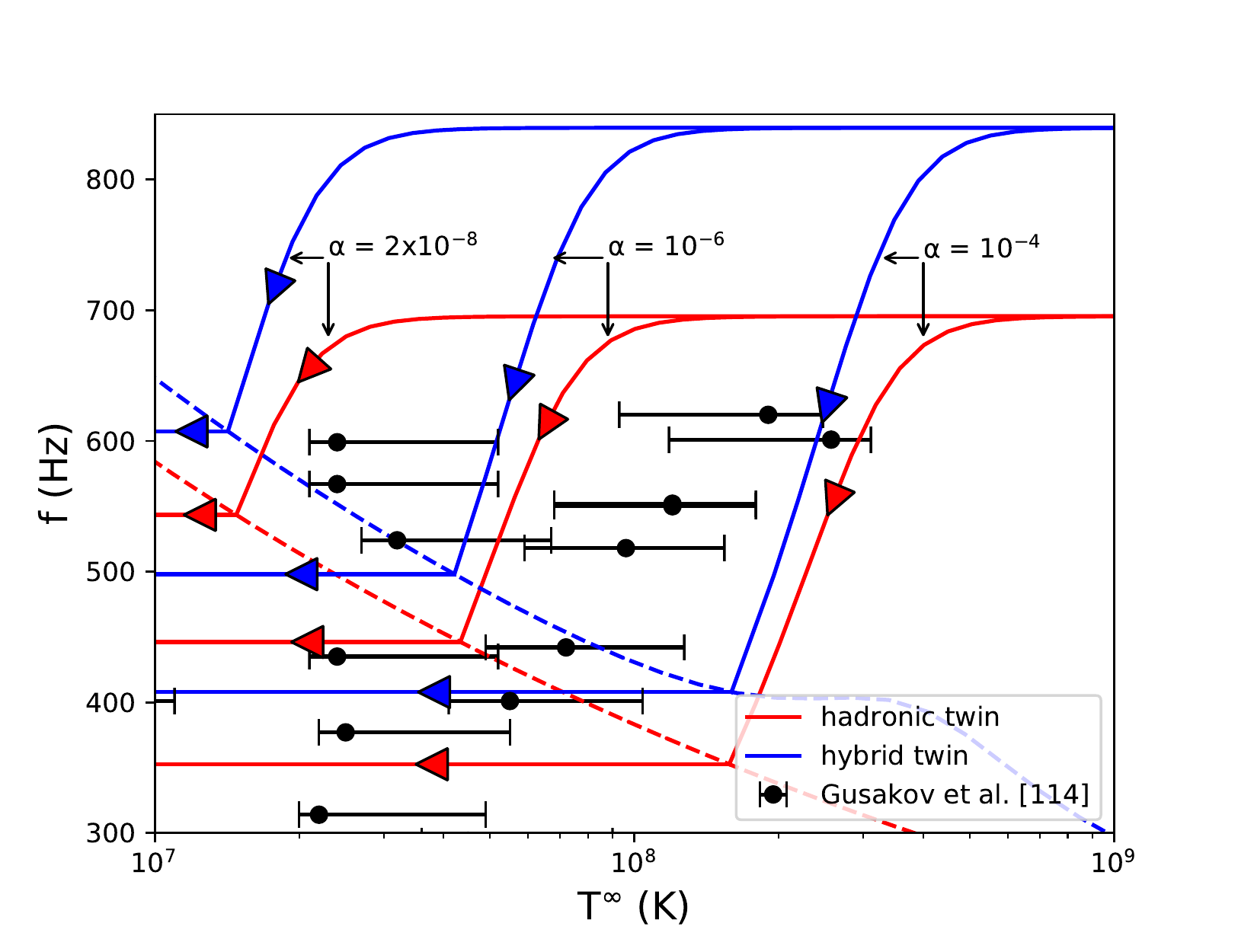}
  \caption{The spin-down evolution of 1.4 $M_\odot$ twin stars (NL3 EOS, $n_{tr}=$ 0.27 fm$^{-3}$ and $\Delta\mathcal{E}=\Delta\mathcal{E}_{cr}+$ 100 MeV fm$^{-3}$) in the frequency-temperature plane for different values of the saturation amplitude.  The initial frequencies for the twins are their corresponding Kepler frequencies. The blue (red) solid lines indicate the evolution for the hybrid (hadronic) twin. The blue and red dashed lines denote the $r$-mode instability window of the hybrid and hadronic star, respectively. The dotted points stand for observational data taken from Ref.~\cite{Gusakov-2014}.}
  \label{spin-down-2} 
\end{figure}

Another observation that can be made from Fig.~\ref{Fig6} is that right after the phase transition occurs a narrowing of the instability window is evident. Then, as the mass further increases the instability window becomes wider for low temperature values ($T^\infty \leq$ 10$^8$ K). The fact that higher mass configurations have wider instability windows is a known result from previous studies \cite{Mukhopadhyay-2018}. It is interesting that while a higher mass is necessary for the stabilization of $r$-modes in observed stars with $T^\infty \geq$ 10$^8$ K, it fails to provide an explanation for the stars appearing in a lower temperature regime. However, the low temperature region can be covered by hybrid star configurations of lower mass. In principle, if a star slightly surpasses a critical mass, after which a phase transition occurs, then its instability window will be also slightly different from the one of the most massive purely hadronic configuration. In contrast, if the structure of the phase transition predicts the existence of a third family, then stars with mass equal or slightly larger than the aforementioned critical mass are going to exhibit considerable deviations in their $r$-mode instability windows. The nontrivial behavior of stars having narrower instability windows compared to those of lower mass stars (for low $T^\infty$), is characteristic of an EOS predicting twin configurations.

\subsection{Spin down and thermal evolution}  

In Fig.~\ref{spin-down-1} we display  the time evolution for the 
frequency and the corresponding spin-down rate of 1.4~$M_\odot$ twin stars. The upper and lower panels contain results for the DD2 and NL3 EOSs, respectively. For comparison reasons, we consider the same initial frequency of 650~Hz for both twins. Furthermore, in accordance to previous studies \cite{Moustakidis-2015,Mukhopadhyay-2018}, the selected value for the $r$-mode saturation amplitude is $a=$ 2$\times$10$^{-7}$. As it is evident from Fig.~\ref{spin-down-1}, the spin-down rate is slower in hybrid stars,  right after their birth. Specifically, the higher the energy density gap the lower the rate. However, after a certain amount of time the spin-down rates of twin stars converge to the same value. The latter is reflected on the distinct time evolution of the frequency for the two cases. In particular, a hybrid star retains its initial rotational frequency for a longer period of time compared to its hadronic twin.

Usually, it is more convenient to study the spin down evolution of a compact star on the $f-T^\infty$ plane~\cite{Routray-2021}. The latter demands the simultaneous knowledge of the spin and thermal evolution for a star. By employing the toy model for the fall of temperature, presented in Sec.~\ref{4}, we intend to obtain the different evolution paths of twin stars on the $f-T^\infty$ plane. In addition, instead of considering the same initial frequency for the two twins we set as initial condition the corresponding Kepler frequencies. The results presented in Fig.~\ref{spin-down-2} were constructed using the NL3 EOS with $n_{tr}=$ 0.27 fm$^{-3}$ (hence 1.4~$M_\odot$ twins) and $\Delta\mathcal{E}=\Delta\mathcal{E}_{cr}+$ 100 MeV fm$^{-3}$. The latter EOS predicts twins with a $\sim$ 13 $\%$ difference in compactness, and therefore, a similar thermal evolution is not an unreasonable assumption \cite{Lyra-2023}. For the crust elasticity a low value of $\mathcal{S}=$ 0.1 was chosen. Moreover, we consider three different values of the amplitude $\alpha$ since the results are very sensitive to it. From   Fig.~\ref{spin-down-2} it is obvious that there are three main reasons which differentiate the time evolution of the two branches. The first one is the different Kepler velocities. The second one is connected to the spin down rates of the two twins, even though this effect is less pronounced. The third one is the deviation of the instability windows. In particular, the unstable region is more extended in the case of the hadronic branch. The latter is of most importance, as the $r$-mode instability window essentially sets the resulting frequency of a star as it comes out of the unstable region. Of course we need to stress out that, the paths presented in  Fig.~\ref{spin-down-2} can  be  improved if one considers a  more realistic cooling process for the two branches. However,  the general picture  will not change noticeably and the main conclusions of the present study are not expected to be significantly altered.

\section{Conclusion} \label{6}

The present work was dedicated to the study of twin stars and their $r$-mode instability windows. In particular, we have conducted a detailed investigation of the parameters that affect the deviation between the instability windows of twin stars. This is of most importance as two stars with identical mass may have different rotational limits. More precisely, two stars in the same region of the frequency-temperature diagram may behave differently with respect to $r$-modes. Subsequently, the future detection of ($r$-mode) GW emission, from stars that are considered to be stable with respect to $r$-modes (due to existing observations), would be a clear sign for the existence of a third family and hence of hadron-quark phase transition.

Firstly, we studied the influence of the energy density jump $\Delta\mathcal{E}$ on the deviation between the instability windows of twin stars. We found that $\Delta\mathcal{E}$ regulates the radius difference between twin configurations. In addition, hybrid stars predicted from EOSs with higher $\Delta\mathcal{E}$ exhibit larger quark core fractions. Thus, the differences in the critical spin frequencies of twins become more pronounced as the energy density jump increases. Secondly, we took into consideration the strong and common (for both twins) dissipation mechanism due to the presence of a viscous boundary layer. What we found is that, the characteristic peak appearing in the $r$-mode instability windows of hybrid stars (around $T^\infty$ $\sim$ 3 $\times$ 10$^8$ K) flattens as the crust elasticity increases. However, depending on the selected value of $\Delta\mathcal{E}$, considerable differences in the limiting frequency of the two twins may remain.

Furthermore, we examined if the EOSs constructed in this study (i.e. EOSs predicting a third family of compact objects) are a viable option for the explanation of current LMXBs data. We found that depending on the phase transition onset (transition density) and also the masses of stars in LMXBs, our EOSs may be compatible with the existing observational data. In particular, for EOSs that predict twin stars with 1.2 $M_\odot$, the bulk viscosity of quark matter
is adequate to stabilize $r$-modes for moderately massive stars ($M \leq 1.6~M_\odot$) in the whole $f-T^\infty$
region occupied by the observed stars in LMXBs. As the critical compact star mass for the phase transition to occur increases, more massive configurations are needed for the stabilization of $r$-modes.

Finally, we studied the differences that manifest in the spin-down evolution of twin pairs. We found that the hybrid star retains its initial spin frequency for a larger period of time and this is because its spin-down rate is lower compared to its hadronic twin. Furthermore, we noticed that larger $\Delta\mathcal{E}$ values result into lower spin-down rates for hybrid stars. In addition, by employing a simplified cooling model we evaluated the evolution paths of twins stars on the $f-T^\infty$ plane. The resulting path differences derive from: a) the fact that the Kepler frequencies (initial conditions) of twin stars are different, b) the different spin evolution, which is dependent on the bulk properties of a star, c) the different instability windows of twin stars which essentially control when and with what frequency a star is going to pass in the $r$-mode stable region.

 There are some other issues that a more elaborate study should take into account such as additional damping mechanisms or a more rigorous treatment of the thermal evolution. In addition, it would be interesting to explore the effects of a mixed phase (EOSs constructed with the Gibbs method). Even though such a study (already in progress) would be more complete from a quantitative point of view, we do not expect that our main conclusions will be significantly altered.
 Finally, we need to highlight that even though there are a few studies focusing on the $r$-mode instability and hybrid stars~\cite{Alford-2010,Jaikumar-2008,Alford-2014,Pan-2006,Zheng-2006}, this is the first work dealing with the possible existence of two stars with identical mass and different $r$-mode instability windows. The future detection of GW  associated with unstable $r$-modes may finally allow us to distinguish twin stars.

\section*{Acknowledgements}
The authors would like to thank Professor K. Kokkotas for his useful insight and comments and mainly for the fruitful discussion about the effect of rapid rotation on the $r$-mode spectrum. 

\end{document}